\documentclass[aps,prl,twocolumn,showpacs,nofootinbib,floatfix]{revtex4-1} 

\usepackage[hiresbb]{graphicx}
\usepackage{color} 

\begin{document}


\title{Instability onset and scaling laws of an autooscillating turbulent flow\\in
a complex (dusty) plasma}

\author{M. Schwabe}
\email[]{mierk.schwabe@dlr.de}
\author{S. Zhdanov}
\author{C. R\"ath}
\affiliation{{Institut f\"ur Materialphysik im Weltraum, Deutsches Zentrum
  f\"ur Luft- und Raumfahrt (DLR), 82234 We\ss ling, Germany}}

\date{\today}

\begin{abstract}
We study a complex plasma under microgravity conditions that is first stabilized with an oscillating electric field. Once the stabilization is stopped, the so-called heartbeat instability develops. We study how the kinetic energy spectrum changes during and after the onset of the instability and compare with the double cascade predicted by Kraichnan and Leith for two-dimensional turbulence. The onset of the instability manifests clearly in the ratio of the reduced rates of cascade of energy and enstrophy and in the power-law exponents of the energy spectra. 
\end{abstract}


\maketitle

\paragraph*{Introduction} Turbulence is often cited as one of the main challenges of modern theoretical physics \cite{Bratanov2015}, despite being a long-standing subject of study. Many turbulent systems are more complex than Navier-Stokes flow in a fluid - involving, for instance, complicated ways of energy injection, transfer and dissipation \cite{Bratanov2015}. 
Turbulent pulsations or its analogues occur in systems as diverse as soap films \cite{Shakeel2007}, insect flight \cite{Wang2000a}, lattices of anharmonic
oscillators \cite{Peyrard2002,Daumont2003}, Bose-Einstein condensates \cite{Reeves2012}, and microswimmers \cite{Tierno2008}. The development or, inversely, the decay of turbulence is of special interest and can be studied, for instance, in flames \cite{Haq2006} or behind grids \cite{Mohamed1990}.

In general, when energy is put into a two-dimensional turbulent system, the energy spectrum splits into the inverse energy and direct enstrophy ranges -- the so-called double cascade develops \cite{Kraichnan1967,Leith1968}. Its presence has been supported by computer simulations, but not unambiguously \cite{Rutgers1998,Frisch1995}. Only a few numerical simulations \cite{Boffetta2007,Boffetta2010a,Xia2014} and experiments \cite{Rutgers1998,Bruneau2005,Kameke2011} were able to simultaneously observe both cascades, which is challenging due to the large range of scales necessary to cover both the inverse and direct cascade \cite{Boffetta2012}. Recently, it was suggested that the inverse cascade might not be robust, but depend on friction \cite{Scott2007,Xia2014}. The evolution of the spectrum during the onset of two-dimensional forced turbulence was investigated in \cite{Paret1997}. A recent topic of interest is the transition of weak wave turbulence to wave turbulence with intermittent collapses \cite{Rumpf2015} and intermittency, i.e. strong non-Gaussian velocity fluctuations, in wave turbulence \cite{Falcon2007,Falcon2010,Falcon2010a,Falcon2010b}{, and so-called Janus spectra which differ in streamwise and transverse direction \cite{Liu2016a}}.

In this paper, we present the first study of developing turbulence in dusty/complex plasmas. Complex plasmas consist of microparticles embedded in a low-temperature plasma. The microparticles acquire high charges and interact with each other. They can be visualized individually and thus enable observations on the kinetic level of, for instance, vortices \cite{Akdim2003,Bockwoldt2014,Schwabe2014}, tunneling \cite{Morfill2006}, and channeling \cite{Du2014}. Gravity is a major force acting on the microparticles in ground-based experiments. Under its influence, the particles are located close to the sheath region of the plasma, where strong electric fields compensate for gravity, and strong ion fluxes are present. It is desirable to perform experiments with microparticles under microgravity conditions. Then, the microparticles are suspended in the more homogeneous plasma bulk with a weak electric field \cite{Goree1999}, and the strong ion flux effects such as wake formation \cite{Kompaneets2016a} are avoided. The data presented in this paper were measured using the PK-3 Plus Laboratory \cite{Thomas2008} in the microgravity environment on board the International Space Station. Here, large, symmetric microparticle clouds form in the plasma bulk which typically contain a small central, particle-free void caused by the interplay between the ion drag and electric forces acting on the microparticles \cite{Goree1999}. 

Several authors have recently begun to study turbulence in complex plasmas \cite{Tsai2012,Schwabe2014,Gupta2014,Zhdanov2015}. Using complex plasmas to study turbulence has the advantage that the particles that carry the interaction can be visualized individually,  in contrast to traditional experiments on turbulence \cite{Arneodo2008,Monchaux2012,Kameke2011}, in which the use of tracer particles might not be reliable \cite{Mathai2016}. Turbulence in complex plasmas occurs at low Reynolds numbers, which is common in viscoelastic fluids \cite{Groisman2000}. Furthermore, complex plasmas are usually highly dissipative. Therefore, it is advantageous to study forced turbulence. A good mechanism to induce turbulent pulsations is the self-sustained heartbeat oscillation \cite{Zhdanov2010,Zhdanov2015}. This type of instability is characterized by a regularly pulsating (auto-oscillating) microparticle cloud \cite{Goree1999,Heidemann2011,Pustylnik2012a}. The heartbeat-induced auto-oscillations pump energy into the microparticle cloud and are able to induce turbulence and effectively channel the flow in two dimensions \cite{Zhdanov2015}. In the experiment presented in this letter, we use a developing heartbeat instability and auto-oscillations to study the onset of turbulence and, specifically, the development of the kinetic energy spectrum. In addition to the channeling due to the heartbeat, we limit our analysis to particles that move within a plane for 0.2~s or longer, thereby excluding particles with a significant transverse velocity component.

\begin{figure}
	\includegraphics[width=0.8\linewidth]{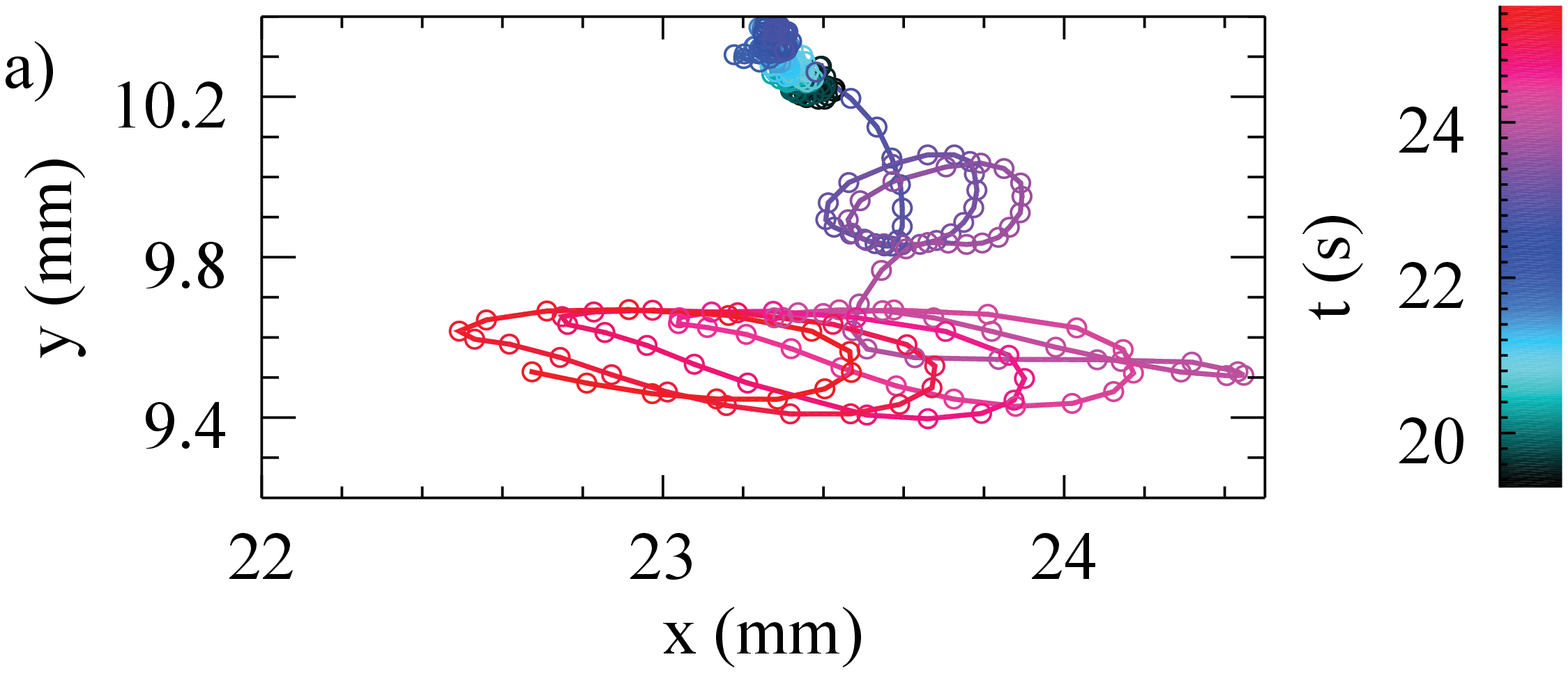}
	\includegraphics[width=0.8\linewidth]{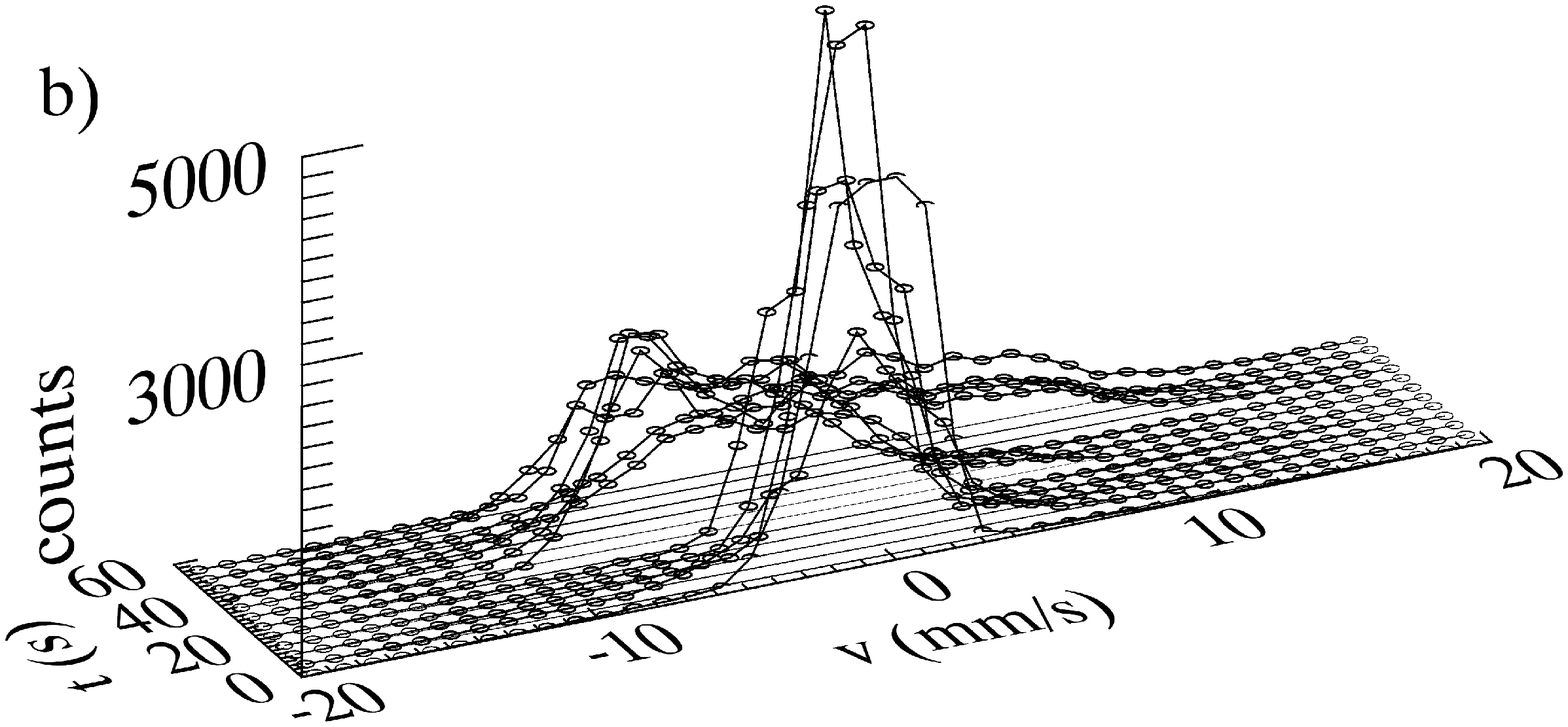}
	\includegraphics[width=0.8\linewidth]{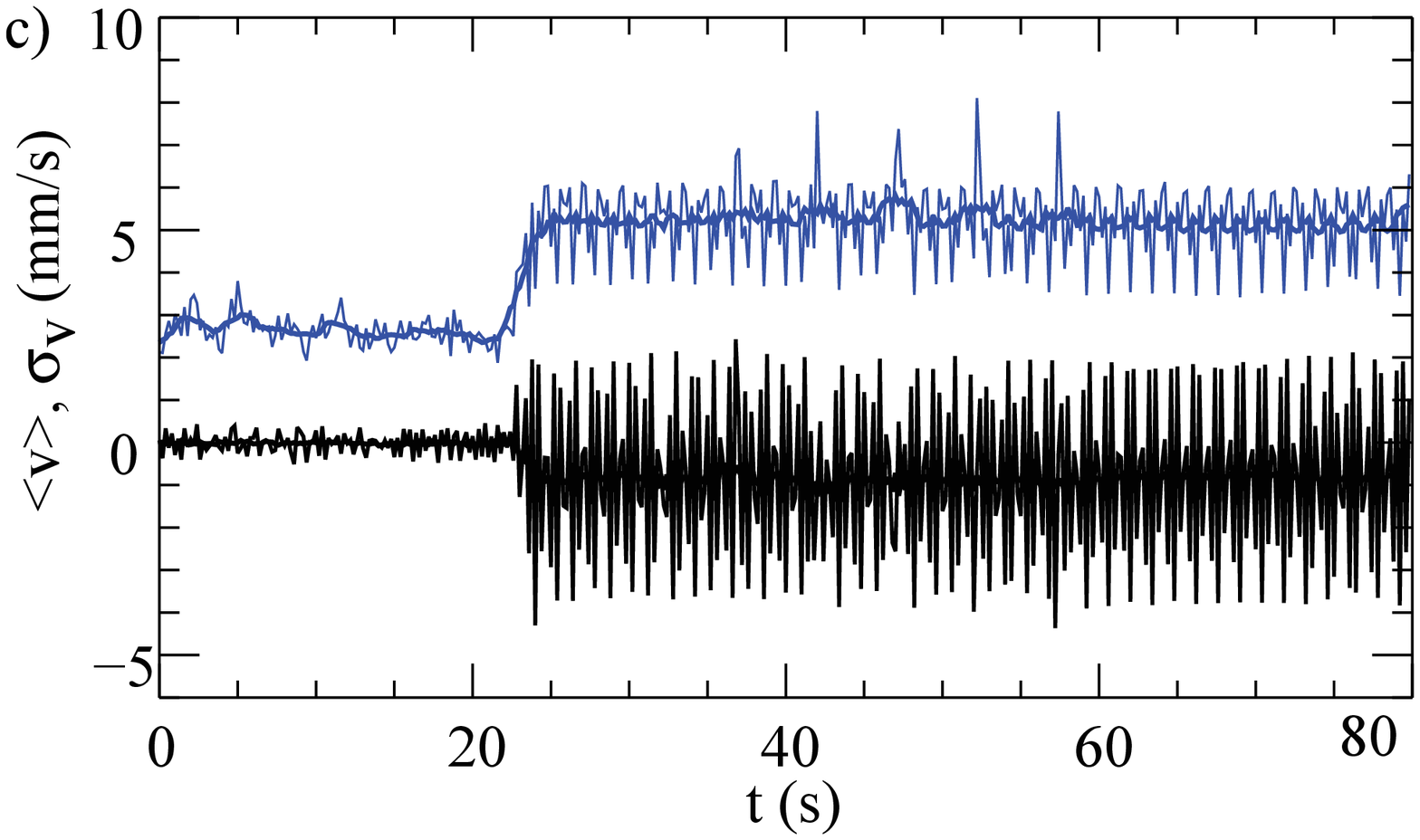}
	\caption{\label{fig:veldist}(color online) a) Selected representative particle track demonstrating alternating dynamics features before, during and after the onset of the instability. Color codes time from $t = 19.3$~s (black) to $t = 25.5$~s (red). b)~Distribution of horizontal velocity {of all particles in the field of view} during the first 60~s of the experiment.  c) Mean horizontal velocity $\left < v \right >$ (bottom line, black) calculated with a sliding window of 0.2~s width, and its standard deviation, $\sigma_v = [ 1/(N-1)\sum_j (v_j - \left < v \right >)^2 ] ^{1/2}$ (top line, blue); $N$ is the number of particles and $v_j$ the velocity of individual particles. When the instability develops ($t = 23$~s), the particles begin to move to the left with  $\left < v \right > \approx -1$~mm/s, and $\sigma_v \approx 5.5$~mm/s. This drift is accompanied by a horizontal oscillation at the heartbeat frequency. This is also visible in the particle track (a), where at $t > 23$~s, the particle starts to move in ovals which are successively displaced leftwards.}
\end{figure}

\paragraph*{Experiment details}Here we present data obtained with the PK-3 Plus Laboratory on board the International Space Station \cite{Thomas2008}. The heart of the laboratory consists of a parallel plate plasma reactor. The experiment was performed in argon at a pressure of 9~Pa. Melamine-formaldehyde particles with a diameter of $(9.2 \pm 1 \%) \, \mu$m and a mass density of 1510~kg/m$^3$ are inserted into the plasma via dispensers. They are illuminated with the light from a laser that is spread into a vertical plane, and their positions are recorded at a frame rate of 50~fps. The rate of damping caused by the friction between microparticles and neutral gas is $\gamma_{damp} = 10.7$~s$^{-1}$ \cite{Epstein1924,Zhdanov2015}. The microparticles' velocity of sound is $C_{DAW} = $ 6--7 mm/s under these conditions~\cite{Zhdanov2015}. 

At the beginning of the experiment presented here, the particle positions are stabilized by applying a rapidly alternating additional electric field (frequency between 42 and 51~Hz). This electric field suppresses \textcolor{black}{a heartbeat} instability. When it is switched off, first the particles move out of the cloud center, and a void develops. This void then begins to regularily expand and contract - it undergoes the heartbeat instability \cite{Goree1999,Thomas2008,Heidemann2011}. The heartbeat in the present experiment has a frequency of $f_{HB} = 2.81 \pm 0.03$~Hz \cite{Zhdanov2015}. We analyze a time series of 80~s duration. The heartbeat develops at $t = 23$~s ($t = 0$~s corresponds to an arbitrarily chosen point during the stabilization stage); see Fig.~\ref{fig:veldist}. The transition from stabilized cloud to heartbeat takes about 2~s as seen from the video images, which is approximately 20~times longer than the time scale defined by friction. A short movie showing experimental data before, during and after the transition can be found in the supplemental material \cite{Supp}.

\paragraph*{Velocity distribution}Figure~\ref{fig:veldist} shows the evolution of the velocity distribution before and after the stabilization is switched off. During the stabilization stage, the mean horizontal velocity is approximately zero. When the stabilization is turned off (at t = 23~s), an oscillation caused by the heartbeat instability develops, causing a drift of the particles leftwards (towards negative x values), see Fig.~\ref{fig:veldist}. Details on the fully developed instability can be found in \cite{Zhdanov2010,Zhdanov2015}. Here we are interested in the transition from the stabilized complex plasma to the unstable system. This transition is clearly seen in the velocity distribution of the microparticles (Fig.~\ref{fig:veldist}). The particles start drifting to the left edge of the cloud, as can be seen in the displacement of the particle track (Fig.~\ref{fig:veldist}a), in the shift of the distribution towards negative values (Fig.~\ref{fig:veldist}b), and in the nonzero mean axial velocity (Fig.~\ref{fig:veldist}c). The movement of the particles with the heartbeat manifests in an oscillating mean velocity. The standard deviation of the velocity, which is a measure of the particle kinetic temperature, also increases once the instability sets in\footnote{A series of strong short-time excitations around $t = $ 36--57~s, caused by external shock compressions of the particle cloud, are also visible. See \cite{Zhdanov2010} for more details.}. {Images visualizing the particle movement during the instability are given in the supplementary material \cite{SuppFlowfield,SuppEns}.}

\begin{figure}
	\includegraphics[width=\linewidth]{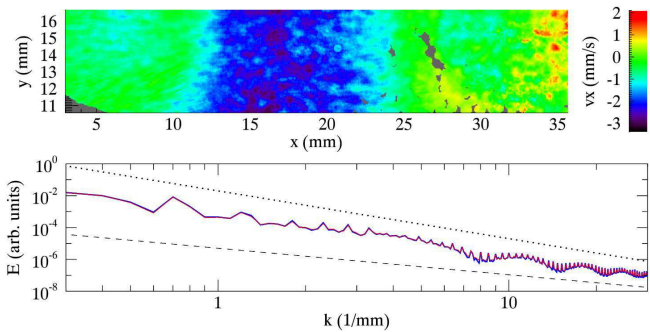}
	\caption{\label{fig:holetest}(color online) Top: Map of the average horizontal velocity during time $t = $ (25--60)~s. Superimposed gray areas indicate the positions of missing data (''holes'') from a different velocity map calculated at time $t= $ (16--21)~s. {The axes indicate the position above the electrode resp. from the left image edge.} Bottom: Energy spectra calculated from the map shown above with no missing data (thick blue line) and from the same map with data removed at the position of the shown holes (thin red line). The two spectra are virtually identical, indicating that the holes superimposed on the map have no effect on the spectrum. The black lines indicate the slopes $E \propto k^{-3}$ (dotted) and $E \propto k^{-5/3}$ (dashed).}
\end{figure}

\paragraph*{Energy spectra}Next, we are going to explore the evolution of the energy spectra of the microparticles. To calculate the spectra, we follow the method presented in \cite{Schwabe2014,Zhdanov2015}: We calculate velocity maps, i.e., the average horizontal or vertical velocity of the particles as a function of position. The energy is then calculated from the squares of the Fourier transformed velocity maps by associating every energy value with the corresponding wave vector modulus $k = |\mathbf{k}|$. As we are interested in the transition process, we calculate the velocity maps averaging over a number of frames. A small number of frames is desirable to increase the resolution during the transition, but introduces the problem of ``holes'' in the velocity maps. These holes are positions in space for which no average velocity data is available, as no particles were present or detected at these positions in the frame range used to calculate the map. The presence of the holes/gaps in the data can potentially have a significant influence on calculated energy spectra \cite{Arevalo2012}.

\begin{figure}
  \centering	
  \includegraphics[width=0.7\linewidth]{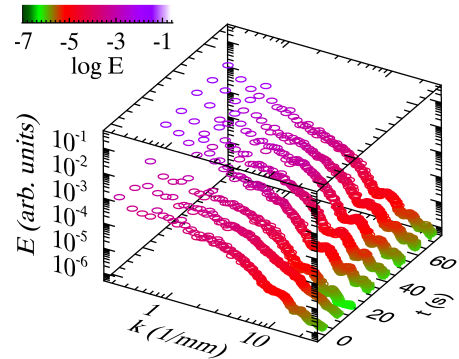}
  \caption{\label{fig:spectra}(color online) Energy spectra with kinetic energy $E$ on the vertical axis as a function of wave number $k$ and time $t$. The color codes the \textcolor{black}{base 10} logarithm of the energy. The rise of the energy when the instability starts is well visible.}
\end{figure}

\begin{figure}
	\includegraphics[width=0.9\linewidth]{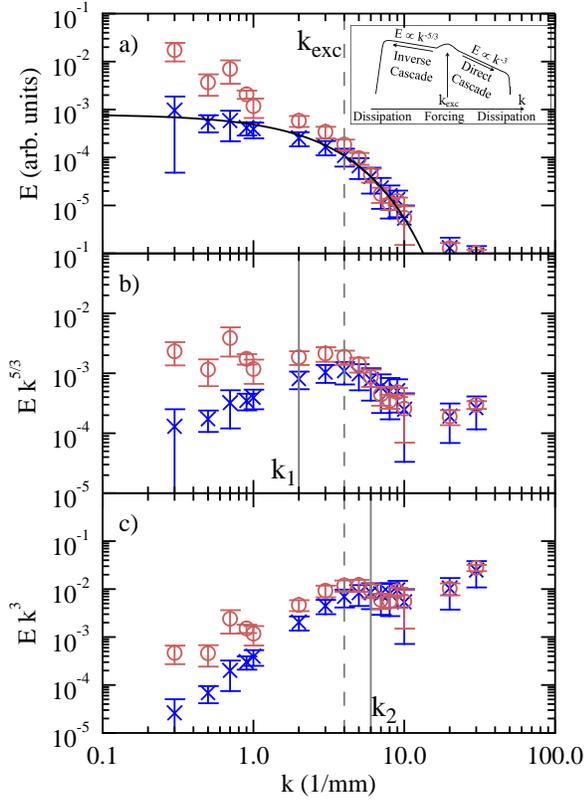} 
	\caption{\label{fig:slopes}(color online) Average energy spectrum calculated before [$t=$(0--22)~s, blue crosses] and after [$t=$(26--76)~s, red circles] the instability starts. a) Energy vs. wave number, overplotted is an exponential dependence (solid line) with $E(k) \propto \exp(-(k/k_0))$ with $k_0 = 2$~mm$^{-1}$. Inset: Schematic demonstrating the double cascade of two-dimensional turbulence, following \cite{Laurie2012}. b),c) Compensated energy spectra. The two solid vertical lines indicate at which wave numbers $\epsilon'$ resp. $\eta'$ were calculated (comp. Fig.~\ref{fig:etaeps}). The dashed line indicates the wave number at which the two ranges meet, $k_{exc}$.}
\end{figure}

\begin{figure}
	\includegraphics[width=\linewidth]{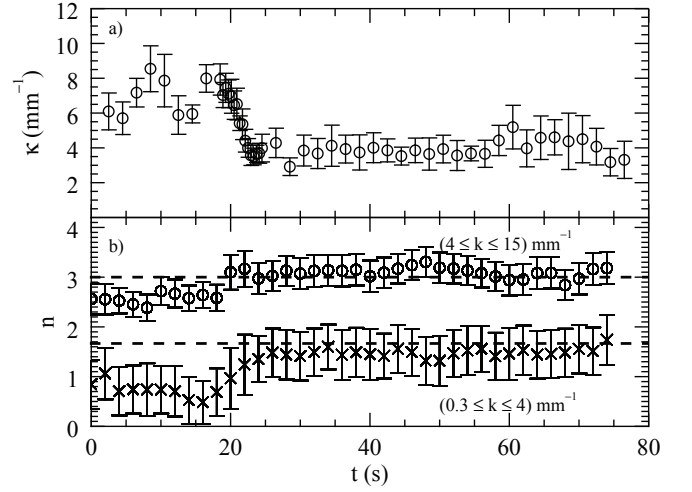}
	\caption{\label{fig:etaeps}{a)} Parameter $\kappa = \sqrt{\eta' / \epsilon'}$  [Eq.~(\ref{eq:kappa})] as a function of time. The onset of the instability at $t = 23$~s is clearly visible as a strong decrease of $\kappa$ during approximately 3~s. {b) Power-law exponents $n$ obtained from linear fits to $\log E = a - n \log k$ as a function of time, where $a$ and $n$ are fit parameters. Fits were done for wave numbers in the range $(0.3 \leq k \leq 4)$~mm$^{-1}$ (bottom) and $(4 \leq k \leq 15)$~mm$^{-1}$ (top), which correspond to the inverse resp. direct cascade ranges, comp. Fig.~\ref{fig:slopes}. The error bars correspond to the 1$\sigma$ uncertainty estimates from the fit. The horizontal dashed lines indicate the values $n = 3$ (top) resp. $n=5/3$ (bottom), which are the exponents predicted for the two cascades in 2d forced turbulence.}}
\end{figure}

The size and number of holes depends on the radius over which mean velocities are calculated and on the number of frames involved in averaging. We test the influence on the spectra by first calculating the spectrum from a map without holes. Then, we artificially remove the data at the position of the holes in a different velocity map and recalculate the energy spectrum. We repeat this method using various averaging radii and frame ranges to find the optimal parameters. According to this analysis, we determine an averaging radius of 5~pixels and a sliding window of 5~s width\footnote{\textcolor{black}{The averaging time is smaller than, for example, the large eddy turnover time and the oscillon period $\tau_{LE} \approx \tau_{osc} \approx 10$~s.}} as optimal choice to calculate velocity maps. The result is shown in Fig.~\ref{fig:holetest}, which clearly demonstrates that there is no significant influence of the holes for the selected parameters. These are thus the parameters selected to calculate the velocity maps used in the following analysis, especially Fig.~\ref{fig:spectra}, \ref{fig:slopes}, and \ref{fig:etaeps}.

\paragraph*{Evolution of energy spectra}Figure~\ref{fig:spectra} shows energy spectra calculated in the stages before, during, and after the heartbeat instability sets in. The energy rise once the instability begins is clearly seen, as is the change in slope of the spectra. We plot the average energy spectra before and after the onset of the instability in Figure~\ref{fig:slopes}. Before the onset of the instability, the spectrum (blue crosses in Fig.~\ref{fig:slopes}) shows an exponential dependence. During the instability (red circles in Fig.~\ref{fig:slopes}), the slope of the spectrum changes at small wave numbers, and a range with $E \propto k^{-5/3}$ develops. This is more easily visible in the compensated energy spectra Fig.~\ref{fig:slopes}b) and \ref{fig:slopes}c), which depict the energy multiplied by $k^{5/3}$ resp. $k^3$. There is an almost constant $E \times k^{5/3}$ range in Fig.~\ref{fig:slopes}b) at $k < 4$~mm$^{-1}$,whereas $E \times k^{3}$ in Fig.~\ref{fig:slopes}c) is almost constant at $k > 4$~mm$^{-1}$. The transition between the two ranges occurs at $k_{exc} \approx 4$~mm$^{-1}$. The enstrophy spectrum is given in the supplementary material \cite{SuppEns}. 
Next, we determine the power-law exponents $n$ of the energy spectra in the two ranges as a function of time and plot them in Figure~\ref{fig:etaeps}b). The transition when the instability sets in is well visible for both ranges (note that the obtained values of the slopes are sensitive to the exact $k$ ranges selected, but the results remain qualitatively unchanged).

\paragraph*{Comparison to 2d forced turbulence}
Our experiment is intrinsically three-dimensional (3d) in nature. Typically, in 3d turbulence, the energy spectrum follows a $E \propto k^{-5/3}$ law over a suitable range \cite{Frisch1995}. It is for a two-dimensional (2d) system into which energy is injected at a length scale $\ell_{exc}$ that \citet{Kraichnan1967} and \citet{Leith1968} predicted a separation of the spectrum into scales larger and smaller than $\ell_{exc}$: For $k < k_{exc} = 2 \pi / \ell_{exc}$, energy is transferred to lower wave numbers $k$ at zero vorticity flow by elongation and thinning of vortices \cite{Chen2006}. For this inverse energy cascade, it holds that 
\begin{equation}
	E = C \epsilon^{2/3} k^{-5/3}, \label{eq:eps}
\end{equation}
where $E$ is the energy and $C$ is a dimensionless positive coefficient. The rate of cascade of kinetic energy per unit mass is signified by $\epsilon$. For $k > k_{exc}$, it holds that 
\begin{equation}
	E = \tilde{C} \eta^{2/3} k^{-3}, \label{eq:eta}
\end{equation}
where $\tilde{C}$ is another positive constant, and $\eta$ is the rate of cascade of mean-square vorticity. In this enstrophy cascade range (the direct cascade), vorticity flows \textcolor{black}{from large to small spatial scales, and there is a minor energy cascade in the same direction \cite{Boffetta2012,Eyink1996}}\footnote{There is a possible logarithmic correction to Eq.~(\ref{eq:eta}) of the form $[\ln(k/k_{exc})]^{-1/3}$ \cite{Kraichnan1971}. The presence of linear friction can lead to steeper power laws than given by Eq.~(\ref{eq:eta}) \cite{Boffetta2012}.}. This power law is also displayed by freely decaying two-dimensional turbulence \cite{Batchelor1969}. The double cascade structure is depicted schematically in the inset of Figure~\ref{fig:slopes}. 

\textcolor{black}{The power-law exponents $n$ obtained from linear fits to the spectra in the present experiment are shown in Fig.~\ref{fig:etaeps}b. The average values for $t \geq 25$~s are $n = 3.1 \pm 0.3$ for $(4 \leq k \leq 15)$~mm$^{-1}$ and $n = 1.5 \pm 0.5$ for $(0.3 \leq k \leq 4)$~mm$^{-1}$. Apparently, these values agree well (within the experimental uncertainties) with Kraichnan's power-law exponents $n=3$ and $n=5/3$.}

Applying Kraichnan and Leith's theory, it is possible to calculate the reduced rates of energy and enstrophy transfer, $\eta` = \tilde{C}^{3/2} \eta$ and $\epsilon` = C^{3/2} \epsilon$ with equations~(\ref{eq:eps}) and (\ref{eq:eta}) using the energy at two wave numbers $k_1$ and $k_2$ that fall into the two ranges. We select $k_1 = 2$~mm$^{-1}$ and $k_2 = 6$~mm$^{-1}$, which are indicated in Figure~\ref{fig:slopes} with vertical grey lines. Next, we define a parameter $\kappa$
\begin{equation}
	\kappa^2 = \eta' / \epsilon' = \left( \frac{E k^{3}\vert_{k_2}}  {E k^{5/3}\vert_{k_1}} \right)  ^{3/2}. \label{eq:kappa}
\end{equation}
Figure~\ref{fig:etaeps}a) shows $\kappa$ determined with Eq.~(\ref{eq:kappa}) as a function of time. As can be seen, $\kappa$ behaves stepwise:  Before the instability begins, $\left< \kappa \right> = (7 \pm 1)$~mm$^{-1}$. Around the onset of the instability, there is a strong decrease of $\kappa$ during approximately 3~s, afterwards $\left< \kappa \right> = (3.9 \pm 0.5)$~mm$^{-1}$. When a double cascade is present, $\kappa$ can be used to determine the excitation wave number $k_{exc}$: It is exactly at this wave number that the direct and indirect ranges are linked. Thus, at $k_{exc}$, Eqs~(\ref{eq:eps}) and (\ref{eq:eta}) can be equated, giving $\epsilon'^{2/3} k_{exc}^{-5/3} = \eta'^{2/3} k_{exc}^{-3}$, and  $k_{exc} = \sqrt{\eta'/\epsilon'} = \kappa$. The value of $k_{exc} \approx 4$~mm$^{-1}$ fits very well with the transition between the two ranges observed in Fig.~\ref{fig:slopes}. Note that this value of $k_{exc}$ agrees with, but is somewhat larger than, the previously estimated value of $k_{exc} = 2 \pi f_{HB} / C_{DAW}\approx 2.7$~mm$^{-1}$~\cite{Zhdanov2015}.  

\paragraph*{Discussion and conclusion}

\textcolor{black}{We speculate that} the energy spectra that we observe indicate a double cascade as predicted by \citet{Kraichnan1967} and \citet{Leith1968} for forced two-dimensional turbulence. The fact that the excitation wave number estimated with the ratio of the reduced energy and enstrophy transfer rates corresponds well to that determined directly from the energy spectrum furthermore supports this speculation. If true, this would be one of the few experiments on turbulence in which both ranges of the double cascade are observed simultaneously. 

\textcolor{black}{However, the coincidence between the scaling exponents from Kraichnan's work and those found in the present work are surprising, as the former exponents were obtained originally under conservation laws for an inviscid 2D flow, and our system is dissipative due to particle-particle interactions. We cannot exclude that the coincidence between the power-law exponents might be caused by other effects such as a flow from the third dimension. A future, more precise investigation is warranted.} A simultaneous simulation of the plasma and microparticle dynamics would be best, but is complicated by the vastly different time scales. Previous simulations on (unforced) two-dimensional turbulence in complex plasmas are promising \cite{Schwabe2014}. Further experiments dedicated to turbulence in complex plasmas could include measuring fluxes, the decay of turbulent motion, and investigating in more detail the microparticle trajectories. 

\begin{acknowledgments}
\paragraph*{Acknowledgments}We would like to thank Dr. Hubertus Thomas for useful discussions and the \mbox{PK-3} Plus Team at DLR Oberpfaffenhofen, Germany, and at JIHT Moscow, Russia, for relinquishing the data to us. PK-3 Plus was funded by DLR/BMWi under the contract Nos. FKZs 50 WM 0203 and 50 WM 1203.
\end{acknowledgments}

\onecolumngrid
\appendix

\section*{Supplemental figures}

\begin{figure}[h]
	\includegraphics[width=\linewidth]{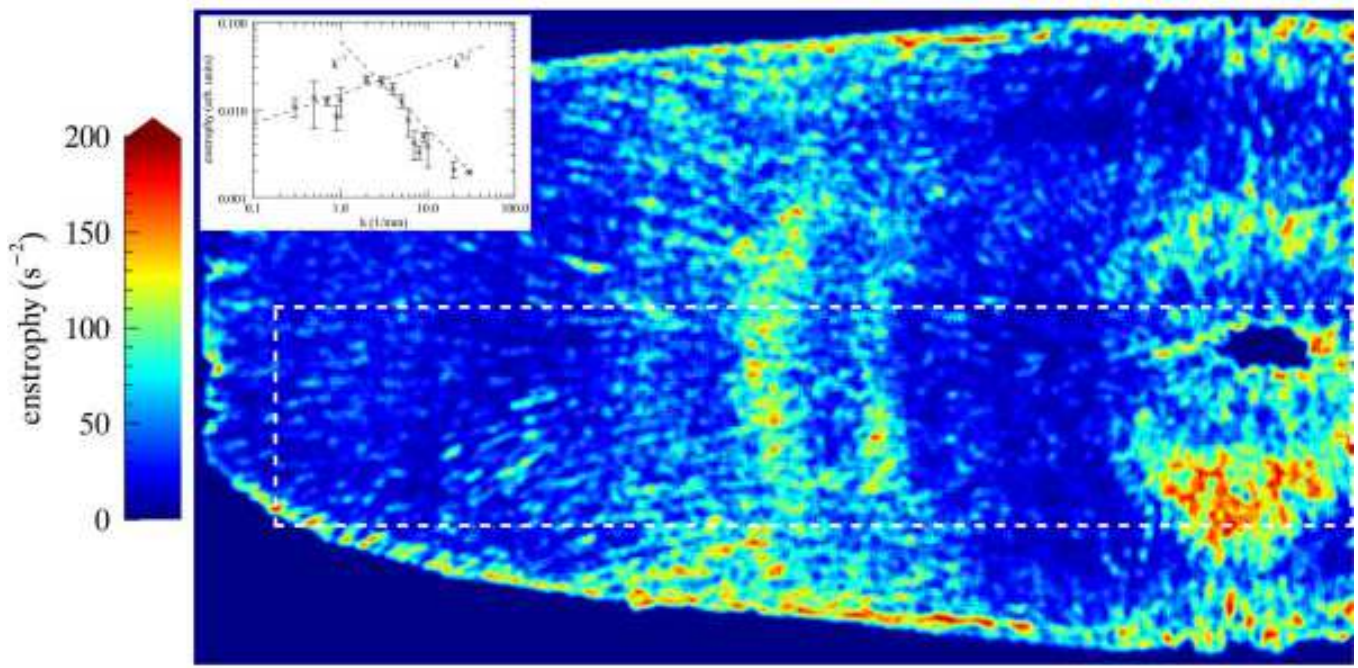}
	\caption{Enstrophy map calculated from particle velocities smoothed spatially with a radius 5 pixel and averaged over the time interval 60 s - 65 s. The field of view is 35.2 $\times$ 18.0 mm$^2$. The color codes enstrophy cut off at a value of 200 s$^-2$. The dashed line indicates the region of interest used to calculate the energy and enstrophy spectra shown in the present figure and in Figs. 2 - 4. This region of interest contains 850 +- 10\% particles at any point of time. (inset) Enstrophy spectrum calculated from the average velocity maps corresponding to the time interval 60 s - 65 s. The dashed lines (shown to guide the eye) are power laws with the exponents -1 and 1/3, which are the spectral exponents corresponding to the 2d enstrophy spectrum of the two-cascades forced turbulence. }
\end{figure}

\begin{figure}[]
	\includegraphics[width=\linewidth]{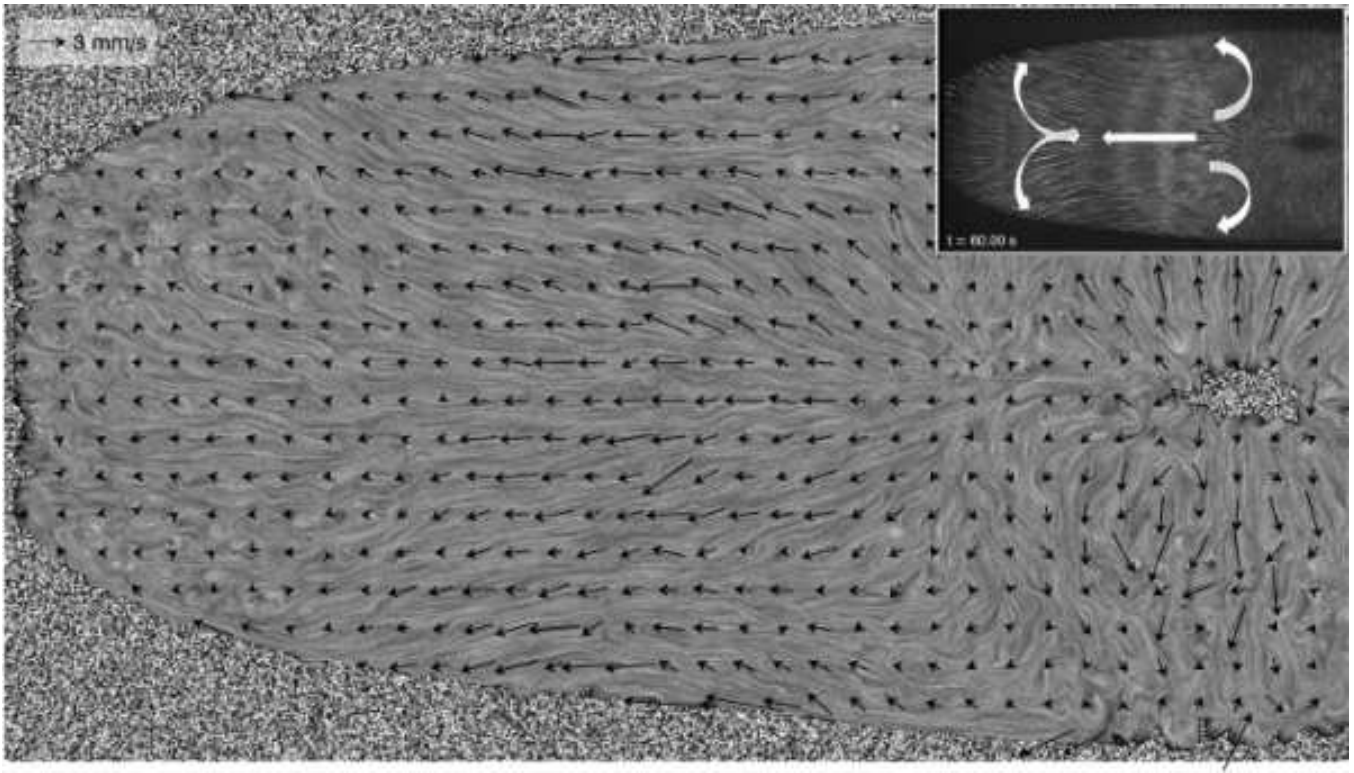}
	\caption{Flow lines of the average velocity field corresponding to time interval 60 s - 65 s. The field of view is 35.2 $\times$ 18.0 mm$^2$. Small vortices in the particle movement are well visible, as is a general drift of particles along the horizontal direction caused by the heartbeat instability. The image was produced by the method of line integral convolution [1], in which a background image with random noise is distorted along the velocity field. The black arrows indicate the velocity field as well. (inset) Superposition of 100 images, corresponding to time interval 60 - 62 s. The field of view is 35.2 $\times$ 19.2 mm$^2$. Small-scale vortices in the particle movement, large-scale quadropol-type global vortex structures (indicated by white arrows), as well as a general axial drift of particles (also indicated by an arrow, directed leftward) are visible. Globally, the drift along the axis is compensated at the left cloud edge by rotations in the large peropherial vortices.\\		
		\newline [1] B. Cabral and L. C. Leedom: "Imaging vector fields using line integral convolution", in Proc. 20th ann. conf. Comp. graph. and interac. techn. (1993)}
\end{figure}


\begin{thebibliography}{60}%
\makeatletter
\providecommand \@ifxundefined [1]{%
 \@ifx{#1\undefined}
}%
\providecommand \@ifnum [1]{%
 \ifnum #1\expandafter \@firstoftwo
 \else \expandafter \@secondoftwo
 \fi
}%
\providecommand \@ifx [1]{%
 \ifx #1\expandafter \@firstoftwo
 \else \expandafter \@secondoftwo
 \fi
}%
\providecommand \natexlab [1]{#1}%
\providecommand \enquote  [1]{``#1''}%
\providecommand \bibnamefont  [1]{#1}%
\providecommand \bibfnamefont [1]{#1}%
\providecommand \citenamefont [1]{#1}%
\providecommand \href@noop [0]{\@secondoftwo}%
\providecommand \href [0]{\begingroup \@sanitize@url \@href}%
\providecommand \@href[1]{\@@startlink{#1}\@@href}%
\providecommand \@@href[1]{\endgroup#1\@@endlink}%
\providecommand \@sanitize@url [0]{\catcode `\\12\catcode `\$12\catcode
  `\&12\catcode `\#12\catcode `\^12\catcode `\_12\catcode `\%12\relax}%
\providecommand \@@startlink[1]{}%
\providecommand \@@endlink[0]{}%
\providecommand \url  [0]{\begingroup\@sanitize@url \@url }%
\providecommand \@url [1]{\endgroup\@href {#1}{\urlprefix }}%
\providecommand \urlprefix  [0]{URL }%
\providecommand \Eprint [0]{\href }%
\providecommand \doibase [0]{http://dx.doi.org/}%
\providecommand \selectlanguage [0]{\@gobble}%
\providecommand \bibinfo  [0]{\@secondoftwo}%
\providecommand \bibfield  [0]{\@secondoftwo}%
\providecommand \translation [1]{[#1]}%
\providecommand \BibitemOpen [0]{}%
\providecommand \bibitemStop [0]{}%
\providecommand \bibitemNoStop [0]{.\EOS\space}%
\providecommand \EOS [0]{\spacefactor3000\relax}%
\providecommand \BibitemShut  [1]{\csname bibitem#1\endcsname}%
\let\auto@bib@innerbib\@empty
\bibitem [{\citenamefont {Bratanov}\ \emph {et~al.}(2015)\citenamefont
  {Bratanov}, \citenamefont {Jenko},\ and\ \citenamefont
  {Frey}}]{Bratanov2015}%
  \BibitemOpen
  \bibfield  {author} {\bibinfo {author} {\bibfnamefont {V.}~\bibnamefont
  {Bratanov}}, \bibinfo {author} {\bibfnamefont {F.}~\bibnamefont {Jenko}}, \
  and\ \bibinfo {author} {\bibfnamefont {E.}~\bibnamefont {Frey}},\ }\href@noop
  {} {\bibfield  {journal} {\bibinfo  {journal} {PNAS}\ }\textbf {\bibinfo
  {volume} {112}},\ \bibinfo {pages} {15048} (\bibinfo {year}
  {2015})}\BibitemShut {NoStop}%
\bibitem [{\citenamefont {Shakeel}\ and\ \citenamefont
  {Vorobieff}(2007)}]{Shakeel2007}%
  \BibitemOpen
  \bibfield  {author} {\bibinfo {author} {\bibfnamefont {T.}~\bibnamefont
  {Shakeel}}\ and\ \bibinfo {author} {\bibfnamefont {P.}~\bibnamefont
  {Vorobieff}},\ }\href {\doibase 10.1007/s00348-007-0334-y} {\bibfield
  {journal} {\bibinfo  {journal} {Exp. Fluids}\ }\textbf {\bibinfo {volume}
  {43}},\ \bibinfo {pages} {125} (\bibinfo {year} {2007})}\BibitemShut
  {NoStop}%
\bibitem [{\citenamefont {Wang}(2000)}]{Wang2000a}%
  \BibitemOpen
  \bibfield  {author} {\bibinfo {author} {\bibfnamefont {Z.~J.}\ \bibnamefont
  {Wang}},\ }\href@noop {} {\bibfield  {journal} {\bibinfo  {journal} {Phys.
  Rev. Lett.}\ }\textbf {\bibinfo {volume} {85}},\ \bibinfo {pages} {2216}
  (\bibinfo {year} {2000})}\BibitemShut {NoStop}%
\bibitem [{\citenamefont {Peyrard}\ and\ \citenamefont
  {Daumont}(2002)}]{Peyrard2002}%
  \BibitemOpen
  \bibfield  {author} {\bibinfo {author} {\bibfnamefont {M.}~\bibnamefont
  {Peyrard}}\ and\ \bibinfo {author} {\bibfnamefont {I.}~\bibnamefont
  {Daumont}},\ }\href@noop {} {\bibfield  {journal} {\bibinfo  {journal}
  {Europhys. Lett.}\ }\textbf {\bibinfo {volume} {59}},\ \bibinfo {pages} {834}
  (\bibinfo {year} {2002})}\BibitemShut {NoStop}%
\bibitem [{\citenamefont {Daumont}\ and\ \citenamefont
  {Peyrard}(2003)}]{Daumont2003}%
  \BibitemOpen
  \bibfield  {author} {\bibinfo {author} {\bibfnamefont {I.}~\bibnamefont
  {Daumont}}\ and\ \bibinfo {author} {\bibfnamefont {M.}~\bibnamefont
  {Peyrard}},\ }\href {\doibase 10.1063/1.1530991} {\bibfield  {journal}
  {\bibinfo  {journal} {Choas: Interd. J. Nonl. Sci.}\ }\textbf {\bibinfo
  {volume} {13}},\ \bibinfo {pages} {624} (\bibinfo {year} {2003})}\BibitemShut
  {NoStop}%
\bibitem [{\citenamefont {Reeves}\ \emph {et~al.}(2012)\citenamefont {Reeves},
  \citenamefont {Anderson},\ and\ \citenamefont {Bradley}}]{Reeves2012}%
  \BibitemOpen
  \bibfield  {author} {\bibinfo {author} {\bibfnamefont {M.~T.}\ \bibnamefont
  {Reeves}}, \bibinfo {author} {\bibfnamefont {B.~P.}\ \bibnamefont
  {Anderson}}, \ and\ \bibinfo {author} {\bibfnamefont {A.~S.}\ \bibnamefont
  {Bradley}},\ }\href {\doibase 10.1103/PhysRevA.86.053621} {\bibfield
  {journal} {\bibinfo  {journal} {Phys. Rev. A}\ }\textbf {\bibinfo {volume}
  {86}},\ \bibinfo {pages} {053621} (\bibinfo {year} {2012})}\BibitemShut
  {NoStop}%
\bibitem [{\citenamefont {Tierno}\ \emph {et~al.}(2008)\citenamefont {Tierno},
  \citenamefont {Golestanian}, \citenamefont {Pagonabarraga},\ and\
  \citenamefont {Sague}}]{Tierno2008}%
  \BibitemOpen
  \bibfield  {author} {\bibinfo {author} {\bibfnamefont {P.}~\bibnamefont
  {Tierno}}, \bibinfo {author} {\bibfnamefont {R.}~\bibnamefont {Golestanian}},
  \bibinfo {author} {\bibfnamefont {I.}~\bibnamefont {Pagonabarraga}}, \ and\
  \bibinfo {author} {\bibfnamefont {F.}~\bibnamefont {Sague}},\ }\href
  {\doibase 10.1021/jp808354n} {\bibfield  {journal} {\bibinfo  {journal} {J.
  Phys. Chem. B}\ }\textbf {\bibinfo {volume} {112}},\ \bibinfo {pages} {16525}
  (\bibinfo {year} {2008})}\BibitemShut {NoStop}%
\bibitem [{\citenamefont {Haq}(2006)}]{Haq2006}%
  \BibitemOpen
  \bibfield  {author} {\bibinfo {author} {\bibfnamefont {M.~Z.}\ \bibnamefont
  {Haq}},\ }\href@noop {} {\bibfield  {journal} {\bibinfo  {journal} {J. Eng.
  Gas Turbines Power}\ }\textbf {\bibinfo {volume} {128}},\ \bibinfo {pages}
  {455} (\bibinfo {year} {2006})}\BibitemShut {NoStop}%
\bibitem [{\citenamefont {Mohamed}\ and\ \citenamefont
  {LaRue}(1990)}]{Mohamed1990}%
  \BibitemOpen
  \bibfield  {author} {\bibinfo {author} {\bibfnamefont {M.~S.}\ \bibnamefont
  {Mohamed}}\ and\ \bibinfo {author} {\bibfnamefont {J.~C.}\ \bibnamefont
  {LaRue}},\ }\href@noop {} {\bibfield  {journal} {\bibinfo  {journal} {J.
  Fluid Mech.}\ }\textbf {\bibinfo {volume} {219}},\ \bibinfo {pages} {195}
  (\bibinfo {year} {1990})}\BibitemShut {NoStop}%
\bibitem [{\citenamefont {Kraichnan}(1967)}]{Kraichnan1967}%
  \BibitemOpen
  \bibfield  {author} {\bibinfo {author} {\bibfnamefont {R.~H.}\ \bibnamefont
  {Kraichnan}},\ }\href {\doibase 10.1063/1.1762301} {\bibfield  {journal}
  {\bibinfo  {journal} {Phys. Fluids}\ }\textbf {\bibinfo {volume} {10}},\
  \bibinfo {pages} {1417} (\bibinfo {year} {1967})}\BibitemShut {NoStop}%
\bibitem [{\citenamefont {Leith}(1968)}]{Leith1968}%
  \BibitemOpen
  \bibfield  {author} {\bibinfo {author} {\bibfnamefont {C.~E.}\ \bibnamefont
  {Leith}},\ }\href {\doibase 10.1063/1.1691968} {\bibfield  {journal}
  {\bibinfo  {journal} {Phys. Fluids}\ }\textbf {\bibinfo {volume} {11}},\
  \bibinfo {pages} {671} (\bibinfo {year} {1968})}\BibitemShut {NoStop}%
\bibitem [{\citenamefont {Rutgers}(1998)}]{Rutgers1998}%
  \BibitemOpen
  \bibfield  {author} {\bibinfo {author} {\bibfnamefont {M.~A.}\ \bibnamefont
  {Rutgers}},\ }\href {\doibase 10.1103/PhysRevLett.81.2244} {\bibfield
  {journal} {\bibinfo  {journal} {Phys. Rev. Lett.}\ }\textbf {\bibinfo
  {volume} {81}},\ \bibinfo {pages} {2244} (\bibinfo {year}
  {1998})}\BibitemShut {NoStop}%
\bibitem [{\citenamefont {Frisch}(1995)}]{Frisch1995}%
  \BibitemOpen
  \bibfield  {author} {\bibinfo {author} {\bibfnamefont {U.}~\bibnamefont
  {Frisch}},\ }\href@noop {} {\emph {\bibinfo {title} {Turbulence : the legacy
  of A.N. Kolmogorov}}}\ (\bibinfo  {publisher} {Cambridge University Press},\
  \bibinfo {year} {1995})\BibitemShut {NoStop}%
\bibitem [{\citenamefont {Boffetta}(2007)}]{Boffetta2007}%
  \BibitemOpen
  \bibfield  {author} {\bibinfo {author} {\bibfnamefont {G.}~\bibnamefont
  {Boffetta}},\ }\href {\doibase 10.1017/S0022112007008014} {\bibfield
  {journal} {\bibinfo  {journal} {J. Fluid Mech.}\ }\textbf {\bibinfo {volume}
  {589}},\ \bibinfo {pages} {253} (\bibinfo {year} {2007})}\BibitemShut
  {NoStop}%
\bibitem [{\citenamefont {Boffetta}\ and\ \citenamefont
  {Musacchio}(2010)}]{Boffetta2010a}%
  \BibitemOpen
  \bibfield  {author} {\bibinfo {author} {\bibfnamefont {G.}~\bibnamefont
  {Boffetta}}\ and\ \bibinfo {author} {\bibfnamefont {S.}~\bibnamefont
  {Musacchio}},\ }\href {\doibase 10.1103/PhysRevE.82.016307} {\bibfield
  {journal} {\bibinfo  {journal} {Phys. Rev. E}\ }\textbf {\bibinfo {volume}
  {82}},\ \bibinfo {pages} {016307} (\bibinfo {year} {2010})}\BibitemShut
  {NoStop}%
\bibitem [{\citenamefont {Xia}\ and\ \citenamefont {Qian}(2014)}]{Xia2014}%
  \BibitemOpen
  \bibfield  {author} {\bibinfo {author} {\bibfnamefont {Y.X.}~\bibnamefont
  {Xia}}\ and\ \bibinfo {author} {\bibfnamefont {Y.H.}~\bibnamefont {Qian}},\
  }\href {\doibase 10.1103/PhysRevE.90.023004} {\bibfield  {journal} {\bibinfo
  {journal} {Phys. Rev. E}\ }\textbf {\bibinfo {volume} {90}},\ \bibinfo
  {pages} {023004} (\bibinfo {year} {2014})}\BibitemShut {NoStop}%
\bibitem [{\citenamefont {Bruneau}\ and\ \citenamefont
  {Kellay}(2005)}]{Bruneau2005}%
  \BibitemOpen
  \bibfield  {author} {\bibinfo {author} {\bibfnamefont {C.~H.}\ \bibnamefont
  {Bruneau}}\ and\ \bibinfo {author} {\bibfnamefont {H.}~\bibnamefont
  {Kellay}},\ }\href {\doibase 10.1103/PhysRevE.71.046305} {\bibfield
  {journal} {\bibinfo  {journal} {Phys. Rev. E}\ }\textbf {\bibinfo {volume}
  {71}},\ \bibinfo {pages} {046305} (\bibinfo {year} {2005})}\BibitemShut
  {NoStop}%
\bibitem [{\citenamefont {von Kameke}\ \emph {et~al.}(2011)\citenamefont {von
  Kameke}, \citenamefont {Huhn}, \citenamefont {Fern\'{a}ndez-Garc\'{i}a},
  \citenamefont {Munuzuri},\ and\ \citenamefont
  {P\'{e}rez-Munuzuri}}]{Kameke2011}%
  \BibitemOpen
  \bibfield  {author} {\bibinfo {author} {\bibfnamefont {A.}~\bibnamefont {von
  Kameke}}, \bibinfo {author} {\bibfnamefont {F.}~\bibnamefont {Huhn}}, et al. }, \href {\doibase
  10.1103/PhysRevLett.107.074502} {\bibfield  {journal} {\bibinfo  {journal}
  {Phys. Rev. Lett.}\ }\textbf {\bibinfo {volume} {107}},\ \bibinfo {pages}
  {074502} (\bibinfo {year} {2011})}\BibitemShut {NoStop}%
\bibitem [{\citenamefont {Boffetta}\ and\ \citenamefont
  {Ecke}(2012)}]{Boffetta2012}%
  \BibitemOpen
  \bibfield  {author} {\bibinfo {author} {\bibfnamefont {G.}~\bibnamefont
  {Boffetta}}\ and\ \bibinfo {author} {\bibfnamefont {R.~E.}\ \bibnamefont
  {Ecke}},\ }\href {\doibase 10.1146/annurev-fluid-120710-101240} {\bibfield
  {journal} {\bibinfo  {journal} {Ann. Rev. Fluid Mech.}\ }\textbf {\bibinfo
  {volume} {44}},\ \bibinfo {pages} {427} (\bibinfo {year} {2012})}\BibitemShut
  {NoStop}%
\bibitem [{\citenamefont {Scott}(2007)}]{Scott2007}%
  \BibitemOpen
  \bibfield  {author} {\bibinfo {author} {\bibfnamefont {R.~K.}\ \bibnamefont
  {Scott}},\ }\href {\doibase 10.1103/PhysRevE.75.046301} {\bibfield  {journal}
  {\bibinfo  {journal} {Phys. Rev. E}\ }\textbf {\bibinfo {volume} {75}},\
  \bibinfo {pages} {046301} (\bibinfo {year} {2007})}\BibitemShut {NoStop}%
\bibitem [{\citenamefont {Paret}\ and\ \citenamefont
  {Tabeling}(1997)}]{Paret1997}%
  \BibitemOpen
  \bibfield  {author} {\bibinfo {author} {\bibfnamefont {J.}~\bibnamefont
  {Paret}}\ and\ \bibinfo {author} {\bibfnamefont {P.}~\bibnamefont
  {Tabeling}},\ }\href {\doibase 10.1103/PhysRevLett.79.4162} {\bibfield
  {journal} {\bibinfo  {journal} {Phys. Rev. Lett.}\ }\textbf {\bibinfo
  {volume} {79}},\ \bibinfo {pages} {4162} (\bibinfo {year}
  {1997})}\BibitemShut {NoStop}%
\bibitem [{\citenamefont {Rumpf}\ and\ \citenamefont
  {Sheffield}(2015)}]{Rumpf2015}%
  \BibitemOpen
  \bibfield  {author} {\bibinfo {author} {\bibfnamefont {B.}~\bibnamefont
  {Rumpf}}\ and\ \bibinfo {author} {\bibfnamefont {T.~Y.}\ \bibnamefont
  {Sheffield}},\ }\href {\doibase 10.1103/PhysRevE.92.022927} {\bibfield
  {journal} {\bibinfo  {journal} {Phys. Rev. E}\ }\textbf {\bibinfo
  {volume} {92}},\ \bibinfo {pages} {022927} (\bibinfo {year} {2015})}\BibitemShut {NoStop}%
\bibitem [{\citenamefont {Falcon}\ \emph {et~al.}(2007)\citenamefont {Falcon},
  \citenamefont {Fauve},\ and\ \citenamefont {Laroche}}]{Falcon2007}%
  \BibitemOpen
  \bibfield  {author} {\bibinfo {author} {\bibfnamefont {E.}~\bibnamefont
  {Falcon}}, \bibinfo {author} {\bibfnamefont {S.}~\bibnamefont {Fauve}}, \
  and\ \bibinfo {author} {\bibfnamefont {C.}~\bibnamefont {Laroche}},\ }\href
  {\doibase 10.1103/PhysRevLett.98.154501} {\bibfield  {journal} {\bibinfo
  {journal} {Phys. Rev. Lett.}\ } \textbf {\bibinfo {volume} {98}},\ \bibinfo {pages} {154501} (\bibinfo {year} {2007})}\BibitemShut {NoStop}%
\bibitem [{\citenamefont {Falcon}(2010)}]{Falcon2010}%
  \BibitemOpen
  \bibfield  {author} {\bibinfo {author} {\bibfnamefont {E.}~\bibnamefont
  {Falcon}},\ }\href {\doibase 10.3934/dcdsb.2010.13.819} {\bibfield  {journal}
  {\bibinfo  {journal} {Disc. Conti. Dyn. Sys. Series B}\ }\textbf {\bibinfo
  {volume} {13}},\ \bibinfo {pages} {819} (\bibinfo {year} {2010})}\BibitemShut
  {NoStop}%
\bibitem [{\citenamefont {Falcon}\ \emph
  {et~al.}(2010{\natexlab{a}})\citenamefont {Falcon}, \citenamefont {Roux},\
  and\ \citenamefont {Laroche}}]{Falcon2010a}%
  \BibitemOpen
  \bibfield  {author} {\bibinfo {author} {\bibfnamefont {E.}~\bibnamefont
  {Falcon}}, \bibinfo {author} {\bibfnamefont {S.~G.}\ \bibnamefont {Roux}}, \
  and\ \bibinfo {author} {\bibfnamefont {C.}~\bibnamefont {Laroche}},\ }\href
  {\doibase 10.1209/0295-5075/90/34005} {\bibfield  {journal} {\bibinfo
  {journal} {EPL}\ }\textbf {\bibinfo {volume} {90}},\ \bibinfo {pages} {34005}
  (\bibinfo {year} {2010}{\natexlab{a}})}\BibitemShut {NoStop}%
\bibitem [{\citenamefont {Falcon}\ \emph
  {et~al.}(2010{\natexlab{b}})\citenamefont {Falcon}, \citenamefont {Roux},\
  and\ \citenamefont {Audit}}]{Falcon2010b}%
  \BibitemOpen
  \bibfield  {author} {\bibinfo {author} {\bibfnamefont {E.}~\bibnamefont
  {Falcon}}, \bibinfo {author} {\bibfnamefont {S.~G.}\ \bibnamefont {Roux}}, \
  and\ \bibinfo {author} {\bibfnamefont {B.}~\bibnamefont {Audit}},\ }\href
  {\doibase 10.1209/0295-5075/90/50007} {\bibfield  {journal} {\bibinfo
  {journal} {EPL}\ }\textbf {\bibinfo {volume} {90}},\ \bibinfo {pages} {50007}
  (\bibinfo {year} {2010}{\natexlab{b}})}\BibitemShut {NoStop}%
\bibitem[]{Liu2016a}
{C.-C. Liu, R.~T. Cerbus, and P. Chakraborty, Phys. Rev. Lett. {\bf 117}, 114502 (2016).}
\bibitem [{\citenamefont {Akdim}\ and\ \citenamefont
  {Goedheer}(2003)}]{Akdim2003}%
  \BibitemOpen
  \bibfield  {author} {\bibinfo {author} {\bibfnamefont {M.~R.}\ \bibnamefont
  {Akdim}}\ and\ \bibinfo {author} {\bibfnamefont {W.~J.}\ \bibnamefont
  {Goedheer}},\ }\href {\doibase 10.1103/PhysRevE.67.056405} {\bibfield
  {journal} {\bibinfo  {journal} {Phys. Rev. E}\ }\textbf {\bibinfo {volume}
  {67}},\ \bibinfo {pages} {056405} (\bibinfo {year} {2003})}\BibitemShut
  {NoStop}%
\bibitem [{\citenamefont {Bockwoldt}\ \emph {et~al.}(2014)\citenamefont
  {Bockwoldt}, \citenamefont {Arp}, \citenamefont {Menzel},\ and\ \citenamefont
  {Piel}}]{Bockwoldt2014}%
  \BibitemOpen
  \bibfield  {author} {\bibinfo {author} {\bibfnamefont {T.}~\bibnamefont
  {Bockwoldt}}, \bibinfo {author} {\bibfnamefont {O.}~\bibnamefont {Arp}},
  \bibinfo {author} {\bibfnamefont {K.~O.}\ \bibnamefont {Menzel}}, \ and\
  \bibinfo {author} {\bibfnamefont {A.}~\bibnamefont {Piel}},\ }\href {\doibase
  10.1063/1.4897181} {\bibfield  {journal} {\bibinfo  {journal} {Phys.
  Plasmas}\ }\textbf {\bibinfo {volume} {21}},\ \bibinfo {pages} {103703}
  (\bibinfo {year} {2014})}\BibitemShut {NoStop}%
\bibitem [{\citenamefont {Schwabe}\ \emph {et~al.}(2014)\citenamefont
  {Schwabe}, \citenamefont {Zhdanov}, \citenamefont {R\"{a}th}, \citenamefont
  {Graves}, \citenamefont {Thomas},\ and\ \citenamefont
  {Morfill}}]{Schwabe2014}%
  \BibitemOpen
  \bibfield  {author} {\bibinfo {author} {\bibfnamefont {M.}~\bibnamefont
  {Schwabe}}, \bibinfo {author} {\bibfnamefont {S.}~\bibnamefont {Zhdanov}},
  et al., }{\bibfield  {journal} {\bibinfo  {journal}
  {Phys. Rev. Lett.}\ }\textbf {\bibinfo {volume} {112}},\ \bibinfo {pages}
  {115002} (\bibinfo {year} {2014})}\BibitemShut {NoStop}%
\bibitem [{\citenamefont {Morfill}\ \emph {et~al.}(2006)\citenamefont
  {Morfill}, \citenamefont {Konopka}, \citenamefont {Kretschmer}, \citenamefont
  {Rubin-Zuzic}, \citenamefont {Thomas}, \citenamefont {Zhdanov},\ and\
  \citenamefont {Tsytovich}}]{Morfill2006}%
  \BibitemOpen
  \bibfield  {author} {\bibinfo {author} {\bibfnamefont {G.~E.}\ \bibnamefont
  {Morfill}}, \bibinfo {author} {\bibfnamefont {U.}~\bibnamefont {Konopka}}, et al.,\ }\href {\doibase
  10.1088/1367-2630/8/1/007} {\bibfield  {journal} {\bibinfo  {journal} {New J.
  Phys.}\ }\textbf {\bibinfo {volume} {8}},\ \bibinfo {pages} {7} (\bibinfo
  {year} {2006})}\BibitemShut {NoStop}%
\bibitem [{\citenamefont {Du}\ \emph {et~al.}(2014)\citenamefont {Du},
  \citenamefont {Nosenko}, \citenamefont {Zhdanov}, \citenamefont {Thomas},\
  and\ \citenamefont {Morfill}}]{Du2014}%
  \BibitemOpen
  \bibfield  {author} {\bibinfo {author} {\bibfnamefont {C.-R.}\ \bibnamefont
  {Du}}, \bibinfo {author} {\bibfnamefont {V.}~\bibnamefont {Nosenko}},
  \bibinfo {author} {\bibfnamefont {S.}~\bibnamefont {Zhdanov}}, \bibinfo
  {author} {\bibfnamefont {H.~M.}\ \bibnamefont {Thomas}}, \ and\ \bibinfo
  {author} {\bibfnamefont {G.~E.}\ \bibnamefont {Morfill}},\ }\href {\doibase
  10.1103/PhysRevE.89.021101} {\bibfield  {journal} {\bibinfo  {journal} {Phys.
  Rev. E}\ }\textbf {\bibinfo {volume} {89}},\ \bibinfo {pages} {021101(R)}
  (\bibinfo {year} {2014})}\BibitemShut {NoStop}%
\bibitem [{\citenamefont {Goree}\ \emph {et~al.}(1999)\citenamefont {Goree},
  \citenamefont {Morfill}, \citenamefont {Tsytovich},\ and\ \citenamefont
  {Vladimirov}}]{Goree1999}%
  \BibitemOpen
  \bibfield  {author} {\bibinfo {author} {\bibfnamefont {J.}~\bibnamefont
  {Goree}}, \bibinfo {author} {\bibfnamefont {G.~E.}\ \bibnamefont {Morfill}},
  \bibinfo {author} {\bibfnamefont {V.~N.}\ \bibnamefont {Tsytovich}}, \ and\
  \bibinfo {author} {\bibfnamefont {S.~V.}\ \bibnamefont {Vladimirov}},\ }\href
  {\doibase 10.1103/PhysRevE.59.7055} {\bibfield  {journal} {\bibinfo
  {journal} {Phys. Rev. E}\ }\textbf {\bibinfo {volume} {59}},\ \bibinfo
  {pages} {7055} (\bibinfo {year} {1999})}\BibitemShut {NoStop}%
\bibitem [{\citenamefont {Kompaneets}\ \emph {et~al.}(2016)\citenamefont
  {Kompaneets}, \citenamefont {Morfill},\ and\ \citenamefont
  {Ivlev}}]{Kompaneets2016a}%
  \BibitemOpen
  \bibfield  {author} {\bibinfo {author} {\bibfnamefont {R.}~\bibnamefont
  {Kompaneets}}, \bibinfo {author} {\bibfnamefont {G.~E.}\ \bibnamefont
  {Morfill}}, \ and\ \bibinfo {author} {\bibfnamefont {A.~V.}\ \bibnamefont
  {Ivlev}},\ }\href {\doibase 10.1103/PhysRevE.93.063201} {\bibfield  {journal}
  {\bibinfo  {journal} {Phys. Rev. E}\ }\textbf {\bibinfo {volume} {93}},\ \bibinfo
  {pages} {063201} (\bibinfo {year} {2016})}\BibitemShut {NoStop}%
\bibitem [{\citenamefont {Thomas}\ \emph {et~al.}(2008)\citenamefont {Thomas},
  \citenamefont {Morfill}, \citenamefont {Fortov}, \citenamefont {Ivlev},
  \citenamefont {Molotkov}, \citenamefont {Lipaev}, \citenamefont {Hagl},
  \citenamefont {Rothermel}, \citenamefont {Khrapak}, \citenamefont
  {S\"{u}tterlin}, \citenamefont {Rubin-Zuzic}, \citenamefont {Petrov},
  \citenamefont {Tokarev},\ and\ \citenamefont {Krikalev}}]{Thomas2008}%
  \BibitemOpen
  \bibfield  {author} {\bibinfo {author} {\bibfnamefont {H.~M.}\ \bibnamefont
  {Thomas}}, \bibinfo {author} {\bibfnamefont {G.~E.}\ \bibnamefont {Morfill}}, et al.,\ }\href {\doibase 10.1088/1367-2630/10/3/033036}
  {\bibfield  {journal} {\bibinfo  {journal} {New J. Phys.}\ }\textbf {\bibinfo
  {volume} {10}},\ \bibinfo {pages} {033036} (\bibinfo {year}
  {2008})}\BibitemShut {NoStop}%
\bibitem [{\citenamefont {Tsai}\ \emph {et~al.}(2012)\citenamefont {Tsai},
  \citenamefont {Chang},\ and\ \citenamefont {I}}]{Tsai2012}%
  \BibitemOpen
  \bibfield  {author} {\bibinfo {author} {\bibfnamefont {Y.-Y.}\ \bibnamefont
  {Tsai}}, \bibinfo {author} {\bibfnamefont {M.-C.}\ \bibnamefont {Chang}}, \
  and\ \bibinfo {author} {\bibfnamefont {L.}\ \bibnamefont {I}},\ }\href
  {\doibase 10.1103/PhysRevE.86.045402} {\bibfield  {journal} {\bibinfo
  {journal} {Phys. Rev. E}\ }\textbf {\bibinfo {volume} {86}},\ \bibinfo
  {pages} {045402} (\bibinfo {year} {2012})}\BibitemShut {NoStop}%
\bibitem [{\citenamefont {Gupta}\ \emph {et~al.}(2014)\citenamefont {Gupta},
  \citenamefont {Ganesh},\ and\ \citenamefont {Joy}}]{Gupta2014}%
  \BibitemOpen
  \bibfield  {author} {\bibinfo {author} {\bibfnamefont {A.}~\bibnamefont
  {Gupta}}, \bibinfo {author} {\bibfnamefont {R.}~\bibnamefont {Ganesh}}, \
  and\ \bibinfo {author} {\bibfnamefont {A.}~\bibnamefont {Joy}},\ }\href
  {\doibase 10.1063/1.4890488} {\bibfield  {journal} {\bibinfo  {journal}
  {Phys. Plasmas}\ }\textbf {\bibinfo {volume} {21}},\ \bibinfo {pages}
  {073707} (\bibinfo {year} {2014})}\BibitemShut {NoStop}%
\bibitem [{\citenamefont {Zhdanov}\ \emph {et~al.}(2015)\citenamefont
  {Zhdanov}, \citenamefont {Schwabe}, \citenamefont {R\"{a}th}, \citenamefont
  {Thomas},\ and\ \citenamefont {Morfill}}]{Zhdanov2015}%
  \BibitemOpen
  \bibfield  {author} {\bibinfo {author} {\bibfnamefont {S.}~\bibnamefont
  {Zhdanov}}, \bibinfo {author} {\bibfnamefont {M.}~\bibnamefont {Schwabe}},
  \bibinfo {author} {\bibfnamefont {C.}~\bibnamefont {R\"{a}th}}, \bibinfo
  {author} {\bibfnamefont {H.~M.}\ \bibnamefont {Thomas}}, \ and\ \bibinfo
  {author} {\bibfnamefont {G.~E.}\ \bibnamefont {Morfill}},\ }\href {\doibase
  10.1209/0295-5075/110/35001} {\bibfield  {journal} {\bibinfo  {journal}
  {EPL}\ }\textbf {\bibinfo {volume} {110}},\ \bibinfo {pages} {35001}
  (\bibinfo {year} {2015})}\BibitemShut {NoStop}%
\bibitem [{\citenamefont {Arn\`{e}odo}\ \emph {et~al.}(2008)\citenamefont
  {Arn\`{e}odo}, \citenamefont {Benzi}, \citenamefont {Berg}, \citenamefont
  {Biferale}, \citenamefont {Bodenschatz}, \citenamefont {Busse}, \citenamefont
  {Calzavarini}, \citenamefont {Castaing}, \citenamefont {Cencini},
  \citenamefont {Chevillard}, \citenamefont {Fisher}, \citenamefont {Grauer},
  \citenamefont {Homann}, \citenamefont {Lamb}, \citenamefont {Lanotte},
  \citenamefont {L\'{e}v\`{e}que}, \citenamefont {L\"{u}thi}, \citenamefont
  {Mann}, \citenamefont {Mordant}, \citenamefont {M\"{u}ller}, \citenamefont
  {Ott}, \citenamefont {Ouellette}, \citenamefont {Pinton}, \citenamefont
  {Pope}, \citenamefont {Roux}, \citenamefont {Toschi}, \citenamefont {Xu},\
  and\ \citenamefont {Yeung}}]{Arneodo2008}%
  \BibitemOpen
  \bibfield  {author} {\bibinfo {author} {\bibfnamefont {A.}~\bibnamefont
  {Arn\`{e}odo}}, \bibinfo {author} {\bibfnamefont {R.}~\bibnamefont {Benzi}}, et al.,\ }\href {\doibase
  10.1103/PhysRevLett.100.254504} {\bibfield  {journal} {\bibinfo  {journal}
  {Phys. Rev. Lett.}\ }\textbf {\bibinfo {volume} {100}},\ \bibinfo {pages}
  {254504} (\bibinfo {year} {2008})}\BibitemShut {NoStop}%
\bibitem [{\citenamefont {Monchaux}(2012)}]{Monchaux2012}%
  \BibitemOpen
  \bibfield  {author} {\bibinfo {author} {\bibfnamefont {R.}~\bibnamefont
  {Monchaux}},\ }\href {\doibase 10.1088/1367-2630/14/9/095013} {\bibfield
  {journal} {\bibinfo  {journal} {New J. Phys.}\ }\textbf {\bibinfo {volume}
  {14}},\ \bibinfo {pages} {095013} (\bibinfo {year} {2012})}\BibitemShut
  {NoStop}%
\bibitem [{\citenamefont {Mathai}\ \emph {et~al.}(2016)\citenamefont {Mathai},
  \citenamefont {Calzavarini}, \citenamefont {Brons}, \citenamefont {Sun},\
  and\ \citenamefont {Lohse}}]{Mathai2016}%
  \BibitemOpen
  \bibfield  {author} {\bibinfo {author} {\bibfnamefont {V.}~\bibnamefont
  {Mathai}}, \bibinfo {author} {\bibfnamefont {E.}~\bibnamefont {Calzavarini}},
  \bibinfo {author} {\bibfnamefont {J.}~\bibnamefont {Brons}}, \bibinfo
  {author} {\bibfnamefont {C.}~\bibnamefont {Sun}}, \ and\ \bibinfo {author}
  {\bibfnamefont {D.}~\bibnamefont {Lohse}},\ }\href {\doibase
  10.1103/PhysRevLett.117.024501} {\bibfield  {journal} {\bibinfo  {journal}
  {Phys. Rev. Lett.}\ }\textbf {\bibinfo {volume} {117}},\ \bibinfo {pages}
  {024501} (\bibinfo {year} {2016})}\BibitemShut {NoStop}%
\bibitem [{\citenamefont {Groisman}\ and\ \citenamefont
  {Steinberg}(2000)}]{Groisman2000}%
  \BibitemOpen
  \bibfield  {author} {\bibinfo {author} {\bibfnamefont {A.}~\bibnamefont
  {Groisman}}\ and\ \bibinfo {author} {\bibfnamefont {V.}~\bibnamefont
  {Steinberg}},\ }\href {\doibase 10.1038/35011019} {\bibfield  {journal}
  {\bibinfo  {journal} {Nature}\ }\textbf {\bibinfo {volume} {405}},\ \bibinfo
  {pages} {53} (\bibinfo {year} {2000})}\BibitemShut {NoStop}%
\bibitem [{\citenamefont {Zhdanov}\ \emph {et~al.}(2010)\citenamefont
  {Zhdanov}, \citenamefont {Schwabe}, \citenamefont {Heidemann}, \citenamefont
  {S\"{u}tterlin}, \citenamefont {Thomas}, \citenamefont {Rubin-Zuzic},
  \citenamefont {Rothermel}, \citenamefont {Hagl}, \citenamefont {Ivlev},
  \citenamefont {Morfill}, \citenamefont {Molotkov}, \citenamefont {Lipaev},
  \citenamefont {Petrov}, \citenamefont {Fortov},\ and\ \citenamefont
  {Reiter}}]{Zhdanov2010}%
  \BibitemOpen
  \bibfield  {author} {\bibinfo {author} {\bibfnamefont {S.~K.}\ \bibnamefont
  {Zhdanov}}, \bibinfo {author} {\bibfnamefont {M.}~\bibnamefont {Schwabe}} et al.,\ }\href {\doibase 10.1088/1367-2630/12/4/043006}
  {\bibfield  {journal} {\bibinfo  {journal} {New J. Phys.}\ }\textbf {\bibinfo
  {volume} {12}},\ \bibinfo {pages} {043006} (\bibinfo {year}
  {2010})}\BibitemShut {NoStop}%
\bibitem [{\citenamefont {Heidemann}\ \emph {et~al.}(2011)\citenamefont
  {Heidemann}, \citenamefont {Cou\"{e}del}, \citenamefont {Zhdanov},
  \citenamefont {S\"{u}tterlin}, \citenamefont {Schwabe}, \citenamefont
  {Thomas}, \citenamefont {Ivlev}, \citenamefont {Hagl}, \citenamefont
  {Morfill}, \citenamefont {Fortov}, \citenamefont {Molotkov}, \citenamefont
  {Petrov}, \citenamefont {Lipaev}, \citenamefont {Tokarev}, \citenamefont
  {Reiter},\ and\ \citenamefont {Vinogradov}}]{Heidemann2011}%
  \BibitemOpen
  \bibfield  {author} {\bibinfo {author} {\bibfnamefont {R.}~\bibnamefont
  {Heidemann}}, \bibinfo {author} {\bibfnamefont {L.}~\bibnamefont
  {Cou\"{e}del}}, et al.,\ }\href {\doibase 10.1063/1.3574905}
  {\bibfield  {journal} {\bibinfo  {journal} {Phys. Plasmas}\ }\textbf
  {\bibinfo {volume} {18}},\ \bibinfo {pages} {053701} (\bibinfo {year}
  {2011})}\BibitemShut {NoStop}%
\bibitem [{\citenamefont {Pustylnik}\ \emph {et~al.}(2012)\citenamefont
  {Pustylnik}, \citenamefont {Ivlev}, \citenamefont {Sadeghi}, \citenamefont
  {Heidemann}, \citenamefont {Mitic}, \citenamefont {Thomas},\ and\
  \citenamefont {Morfill}}]{Pustylnik2012a}%
  \BibitemOpen
  \bibfield  {author} {\bibinfo {author} {\bibfnamefont {M.~Y.}\ \bibnamefont
  {Pustylnik}}, \bibinfo {author} {\bibfnamefont {A.~V.}\ \bibnamefont
  {Ivlev}}, et al.,\ }\href {\doibase
  10.1063/1.4757213} {\bibfield  {journal} {\bibinfo  {journal} {Phys.
  Plasmas}\ }\textbf {\bibinfo {volume} {19}},\ \bibinfo {pages} {103701}
  (\bibinfo {year} {2012})}\BibitemShut {NoStop}%
\bibitem [{\citenamefont {Epstein}(1924)}]{Epstein1924}%
  \BibitemOpen
  \bibfield  {author} {\bibinfo {author} {\bibfnamefont {P.}~\bibnamefont
  {Epstein}},\ }\href@noop {} {\bibfield  {journal} {\bibinfo  {journal} {Phys.
  Rev.}\ }\textbf {\bibinfo {volume} {23}},\ \bibinfo {pages} {710} (\bibinfo
  {year} {1924})}\BibitemShut {NoStop}%
\bibitem []{Supp}%
	See Supplemental Material at PRE for a movie of the microparticle motion.
\bibitem []{SuppFlowfield}%
{See Supplemental Material for a figure visualizing the particle flow.}
\bibitem []{SuppEns}%
{See Supplemental Material for an enstrophy map and spectrum.}
\bibitem [{\citenamefont {Ar\'evalo}\ \emph {et~al.}(2012)\citenamefont
  {Ar\'evalo}, \citenamefont {Churazov}, \citenamefont {Zhuravleva},
  \citenamefont {Hern\'andez-Monteagudo},\ and\ \citenamefont
  {Revnivtsev}}]{Arevalo2012}%
  \BibitemOpen
  \bibfield  {author} {\bibinfo {author} {\bibfnamefont {P.}~\bibnamefont
  {Ar\'evalo}}, \bibinfo {author} {\bibfnamefont {E.}~\bibnamefont {Churazov}},
  et al., }\href
  {\doibase 10.1111/j.1365-2966.2012.21789.x} {\bibfield  {journal} {\bibinfo
  {journal} {Mon. Not. R. Astron. Soc.}\ }\textbf {\bibinfo {volume} {426}},\
  \bibinfo {pages} {1793} (\bibinfo {year} {2012})}\BibitemShut {NoStop}%
\bibitem [{\citenamefont {Chen}\ \emph {et~al.}(2006)\citenamefont {Chen},
  \citenamefont {Ecke}, \citenamefont {Eyink}, \citenamefont {Rivera},
  \citenamefont {Wan},\ and\ \citenamefont {Xiao}}]{Chen2006}%
  \BibitemOpen
  \bibfield  {author} {\bibinfo {author} {\bibfnamefont {S.}~\bibnamefont
  {Chen}}, \bibinfo {author} {\bibfnamefont {R.~E.}\ \bibnamefont {Ecke}},
  \bibinfo {author} {\bibfnamefont {G.~L.}\ \bibnamefont {Eyink}}, \bibinfo
  {author} {\bibfnamefont {M.}~\bibnamefont {Rivera}}, \bibinfo {author}
  {\bibfnamefont {M.}~\bibnamefont {Wan}}, \ and\ \bibinfo {author}
  {\bibfnamefont {Z.}~\bibnamefont {Xiao}},\ }\href {\doibase
  10.1103/PhysRevLett.96.084502} {\bibfield  {journal} {\bibinfo  {journal}
  {Phys. Rev. Lett.}\ }\textbf {\bibinfo {volume} {96}},\ \bibinfo {pages}
  {084502} (\bibinfo {year} {2006})}\BibitemShut {NoStop}%
\textcolor{black}{
	\bibitem [{\citenamefont {Eyink}\ (1996)}]{Eyink1996}%
	\BibitemOpen
	\bibfield  {author} {\bibinfo {author} {\bibfnamefont {G.~L.}\ \bibnamefont
			{Eyink}},\ }\href {\doibase 10.1016/0167-2789(95)00250-2} {\bibfield
		{journal} {\bibinfo  {journal} {Phys. D: Nonlin. Phenomena}\ }\textbf {\bibinfo {volume}
			{91}},\ \bibinfo {pages} {97} (\bibinfo {year} {1996})}}\BibitemShut
{NoStop}%
\bibitem [{\citenamefont {Kraichnan}(1971)}]{Kraichnan1971}%
  \BibitemOpen
  \bibfield  {author} {\bibinfo {author} {\bibfnamefont {R.~H.}\ \bibnamefont
  {Kraichnan}},\ }\href {\doibase 10.1017/S0022112071001216} {\bibfield
  {journal} {\bibinfo  {journal} {J. Fluid Mech.}\ }\textbf {\bibinfo {volume}
  {47}},\ \bibinfo {pages} {525} (\bibinfo {year} {1971})}\BibitemShut
  {NoStop}%
\bibitem [{\citenamefont {Batchelor}(1969)}]{Batchelor1969}%
  \BibitemOpen
  \bibfield  {author} {\bibinfo {author} {\bibfnamefont {G.~K.}\ \bibnamefont
  {Batchelor}},\ }\href@noop {} {\bibfield  {journal} {\bibinfo  {journal}
  {Phys. Fluids}\ }\textbf {\bibinfo {volume} {12}},\ \bibinfo {pages} {II}
  (\bibinfo {year} {1969})}\BibitemShut {NoStop}%
\bibitem [{\citenamefont {Laurie}\ \emph {et~al.}(2012)\citenamefont {Laurie},
  \citenamefont {Bortolozzo}, \citenamefont {Nazarenko},\ and\ \citenamefont
  {Residoni}}]{Laurie2012}%
  \BibitemOpen
  \bibfield  {author} {\bibinfo {author} {\bibfnamefont {J.}~\bibnamefont
  {Laurie}}, \bibinfo {author} {\bibfnamefont {U.}~\bibnamefont {Bortolozzo}},
  \bibinfo {author} {\bibfnamefont {S.}~\bibnamefont {Nazarenko}}, \ and\
  \bibinfo {author} {\bibfnamefont {S.}~\bibnamefont {Residoni}},\ }\href
  {\doibase 10.1016/j.physrep.2012.01.004} {\bibfield  {journal} {\bibinfo
  {journal} {Phys. Rep.}\ }\textbf {\bibinfo {volume} {514}},\ \bibinfo {pages}
  {121} (\bibinfo {year} {2012})}\BibitemShut {NoStop}%
\end{thebibliography}
\end{document}